\documentclass[twocolumn,aps,prb,showpacs,superscriptaddress,floatfix]{revtex4-1}

\usepackage{graphicx}
\usepackage{dcolumn}
\usepackage{bm}
\usepackage{color}
\usepackage{amsmath} 
\newcommand{\angstrom}{\text{\normalfont\AA}}

\begin{document}

\title{Field Induced Magnetic Ordering and Single-ion Anisotropy in the Quasi-1D Haldane Chain Compound SrNi$_{2}$V$_{2}$O$_{8}$: A Single Crystal investigation}

\author{A. K. Bera}
\email{anup.bera@helmholtz-berlin.de}
\affiliation{Helmholtz Zentrum Berlin f{\"u}r Materialien und Energie, D-14109 Berlin, Germany}

\author{B. Lake}
\affiliation{Helmholtz Zentrum Berlin f{\"u}r Materialien und Energie, D-14109 Berlin, Germany}
\affiliation{Institut f{\"u}r Festk{\"o}rperphysik, Technische Universit{\"a}t Berlin, Hardenbergstra{\ss}e 36, D-10623 Berlin, Germany}

\author{A. T. M. N. Islam}
\affiliation{Helmholtz Zentrum Berlin f{\"u}r Materialien und Energie, D-14109 Berlin, Germany}

\author{B. Klemke}
\affiliation{Helmholtz Zentrum Berlin f{\"u}r Materialien und Energie, D-14109 Berlin, Germany}

\author{E. Faulhaber}
\altaffiliation{ Present address: Forschungs-Neutronenquelle Heinz Maier-Leibnitz, Technische Universit{\"a}t M{\"u}nchen, D-85747 Garching, Germany.}
\affiliation{Helmholtz Zentrum Berlin f{\"u}r Materialien und Energie, Gemeinsame Forschergruppe, D-85747 Garching, Germany}

\author{J. M. Law }
\affiliation{Hochfeld Magnetlab Dresden, Helmholtz-Zentrum Dresden-Rossendorf, D-01314 Dresden, Germany }%

\date{\today}

\begin{abstract}
Field-induced magnetic ordering in the Haldane chain compound  SrNi$_{2}$V$_{2}$O$_{8}$ and effect of anisotropy have been investigated using single crystals. Static susceptibility, inelastic neutron scattering, high-field magnetization, and low temperature heat-capacity studies confirm a non-magnetic spin-singlet ground state and a gap between the singlet ground state and triplet excited states. The intra-chain exchange interaction is estimated to be $J \sim 8.9{\pm}$0.1 meV. Splitting of the dispersions into two modes with minimum energies 1.57 and 2.58 meV confirms the existence of single-ion anisotropy $D(S^z){^2}$. The value of  {\it D} is estimated to be $-0.51{\pm}0.01$ meV and the easy axis is found to be along the crystallographic {\it c}-axis. Field-induced magnetic ordering has been found with two critical fields [$\mu_0H_c^{\perp c} = 12.0{\pm}$0.2 T  and $\mu_0H_c^{\parallel c} = 20.8{\pm}$0.5 T at 4.2 K]. Field-induced three-dimensional magnetic ordering above the critical fields is evident from the heat-capacity, susceptibility, and high-field magnetization study. The Phase diagram in the {\it H-T} plane has been obtained from the high-field magnetization. The observed results are discussed in the light of theoretical predictions as well as earlier experimental reports on Haldane chain compounds.
\end{abstract}


\maketitle

\section{\label{sec:intro}Introduction}

Quantum magnets are a special class of magnetic materials in which long-range magnetic order is suppressed or destroyed altogether even at zero temperature by strong quantum spin fluctuations. Spin-1 Heisenberg antiferromagnetic (AFM) chains (also known as Haldane chains) - a special type of quantum magnet  - are of current interest due to their novel magnetic properties.\cite{AffleckJPCM.1.3047} Haldane conjectured that one-dimensional (1D) Heisenberg antiferromagnets with integer-spin have a singlet ground state and gapped magnon ({\it S} = 1) excitations with an energy gap between the singlet ground and triplet first excited states. \cite{HaldanePRL.50.1153,HaldanePLA.93.464} This is in contrast to the gapless continuum of spinon ({\it S} = 1/2) excitations for the half-integer AFM spin-chains.\cite{HaldanePRL.50.1153,HaldanePLA.93.464} Further theoretical study by Sakai and Takahashi for the ground state of the Haldane chain systems predicted a phase diagram in the {\it D}-$J_1$ plane (where {\it D} is the single-ion anisotropy of the form {\it D}({\it S}${^z}$)${^2}$, and $J_1$ is the ratio of the interchain $J_\perp$ to intrachain $J$ exchange coupling).\cite{SakaiPRB.42.4537} The phase diagram suggests the possibility of quantum phase transitions from the spin-liquid state (gapped singlet state) to a three-dimensional (3D) AFM ordered state (Ising-type for $D < 0$ or XY-type for $D > 0$) upon achieving the critical values of the interchain exchange interaction ({\it J}$_{\perp}$) and the single-ion anisotropy ({\it D}). These factors lower the energy of excitations at certain points in reciprocal space and ultimately induce a transition to an AFM ordered state when they exceed threshold values. 

Moreover, for spin-gapped systems in general, application of a magnetic field splits the triplet states due to the Zeeman effect, and one of the triplet states is driven into the ground state at a critical field value $H_c$ at which a transition to a magnetic state occurs. This is a field-induced quantum phase transition (QPT). For $H > H_c$, the ground state is magnetized and the excitations are gapless. Recently, special attention has been paid to the field induced magnetic transition in spin-gap systems including Haldane-chains.\cite{RiceSci.298.760} A number of phenomena have been reported, such as, (i) Bose-Einstein condensation (BEC) of magnons in BaCuSi$_{2}$O$_{6}$,\cite{RueggPRL.98.017202, JaimePRL.93.087203} and ${\mathrm{NiCl}}_{2}\mathrm{\text{-}}4\mathrm{SC}({\mathrm{NH}}_{2}{)}_{2}$ \cite{zapfPRL.96.077204} etc and (ii) a superlattice formation of localized triplets in the frustrated two-dimensional dimer system SrCu$_{2}$(BO$_{3})_{2}$.\cite{KodamaSci.298.395} A BEC of magnons is a $n = 2, z = 2$ QPT (XY symmetry characterized by a quadratic spectrum at $H_c$) which is equivalent to a thermodynamic phase transition in $d+2=5$ dimensions. \cite{sachdevQPT} The applied field acts like a chemical potential and the triplets are populated (by magnons) only above the critical value $H_c$. If the interactions between magnons are absent, the excitations would collapse into a condensate of free bosons known as Bose-Einstein condensation (BEC) that corresponds to the XY AFM ordering in the original spin language.\cite{RueggNature.423.62} On the other hand, if interactions are strong, the bosons could be thought to possess a hard core and so form a gas of free fermions with a Fermi surface. In reality the interactions may be expected to lie midway between these extreme cases, and most gapped magnets are approximately equivalent to an interacting sea of fermions.\cite{KonikPRB.66.144416} Moreover, it was argued that the presence of even weak exchange anisotropy can qualitatively modify the physics of the QPT away from the BEC.\cite{GlazkovPRB.69.184410, KolezhukPRB.70.020403} 

For Haldane chains, it is proposed that there is an interaction between magnons, which fundamentally changes the ground state for $H > H_c$ from a condensate of bosonic magnons to a gapless Luttinger liquid state.\cite{MaedaPRL.99.057205} Presence of magnetic anisotropy and interchain interactions can significantly modify the nature of the field-induced phase transition in Haldane chains, and can lead to more complex behavior and a richer phase diagram which is of current interest for theoretical understanding as well as experimental observation and analysis.

Among the experimentally reported Haldane chain systems, the compound Ni(C$_{2}$H$_{8}$N$_{2}$)$_{2}$NO$_{2}$(ClO$_{4}$) (abbreviated as NENP) with planar anisotropy ($D > 0$), does not show any evidence of field-induced magnetic ordering down to 0.2 K in fields up to 13 T.\cite{RenardEPL.3.945, KobayashiJPSJ.61.1772} Instead, the existence of an energy gap was revealed even at {\it H}$_{c}$ (at the onset of magnetization i.e., 13, 7.5, and 11 T, for the $a$, $b$, and $c$ axes at 1.3 K).\cite{KobayashiJPSJ.61.1772} This result was explained by the presence of a staggered field at the Ni$^{2+}$ positions due to the presence of two crystallographically inequivalent sites for Ni$^{2+}$.\cite{ChibaPRB.44.2838, FujiwaraPRB.47.11860} This staggered field causes a small energy gap near the critical field $H_c$ resulting in, a gradual crossover from the gapped singlet state to a magnetically polarized state, rather than a field induced phase transition.\cite{MitraPRL.72.912, SakaiJPSJ.63.867} On the other hand, a field-induced transition to a magnetic order state has been reported for the Ni(C$_{5}$H$_{14}$N$_{2}$)$_{2}$N$_{3}$(PF$_{6}$) (abbreviated as NDMAP) and Ni(C$_{5}$H$_{14}$N$_{2}$)$_{2}$N$_{3}$(ClO$_{4}$) (abbreviated as NDMAZ) Haldane chain compounds which also have planar anisotropy ($D > 0$).\cite{HondaJPCM.9.L83, HondaPRB.63.064420, KobayashiJPSJ.70.813, ChenPRL.86.1618, ZheludevEPL.55.868} Interestingly, unusual spin excitations such as, the existence of three distinct excitations in the ordered phase were observed by ESR and inelastic neutron scattering.\cite{HagiwaraPRL.91.177601, ZheludevPRB.68.134438} This feature is reasonably different from that in the conventional N{\'e}el state, in which spin-wave modes are the dominant excitations.\cite{ZheludevAPA.74.S1} 

Recently, field-induced magnetic ordering was reported for another Haldane chain compound PbNi$_{2}$V$_{2}$O$_{8}$, with uniaxial-anisotropy ($D < 0$) from a high-field magnetization study on aligned powder samples. Two critical fields of ${\mu}_{0}${\it H}$_{c}^{\perp}$= 14.0 T and ${\mu}_{0}${\it H}$_{c}^{\parallel}$ = 19.0 T, respectively, at 4.2 K were reported for two field directions.\cite{TsujiiPRB.72.104402, UchiyamaPRL.83.632, SmirnovPRB.77.100401} For $H > H_c$, the $M(T)$ curve showed a cusp-like minimum at $T_c$ followed by convex type behavior upon decreasing temperature further, which was explained on the basis of BEC.\cite{TsujiiPRB.72.104402} It was proposed that the magnon BEC picture could be applicable as an approximation for this compound as well as for other Haldane chain systems in general.\cite{TsujiiPRB.72.104402} On the other hand, the ESR spectrum of triplet excitations showed a temperature dependence that was explained on the basis of the interaction between the excitations.\cite{SmirnovPRB.77.100401} Furthermore, the powder neutron scattering study revealed the existence of substantial interchain interactions and anisotropy.\cite{ZheludevPRB.62.8921} 

In the context of field-induced magnetic transitions in the presence of inter-chain interaction and uniaxial anisotropy, the compound SrNi$_{2}$V$_{2}$O$_{8}$ which is isostructural to PbNi$_{2}$V$_{2}$O$_{8}$ is of particular interest. The special features of SrNi$_{2}$V$_{2}$O$_{8}$ are (i) the anisotropy (uniaxial); where the induced splitting of the triplet states is comparable to the energy gap, (ii) the critical inter-chain interactions, as well as (iii) the presence of contradictory reports on the nature of the ground state whether an ordered or spin-liquid state. Unlike PbNi$_{2}$V$_{2}$O$_{8}$, an ordered AFM ground state was proposed by Zheludev {\it et. al}.\cite{ZheludevPRB.62.8921} for SrNi$_{2}$V$_{2}$O$_{8}$, from the powder inelastic neutron scattering study. However, the nuclear magnetic resonance (NMR) study on powder samples of SrNi$_{2}$V$_{2}$O$_{8}$ by Pahari {\it et. al}.\cite{PahariPRB.73.012407} found a non-magnetic spin-liquid (singlet) ground state. Our previous DC-magnetization study on powder samples also showed a spin-liquid type magnetic ground state for SrNi$_{2}$V$_{2}$O$_{8}$.\cite{BeraPRB.86.024408} These inconsistent reports suggest that SrNi$_{2}$V$_{2}$O$_{8}$ is indeed situated close to the phase boundary between spin-liquid (nonmagnetic) and 3D AFM (Ising-N{\'e}el type) ordered states in the Sakai-Takahashi phase diagram.\cite{SakaiPRB.42.4537} 

The presence of substantial inter-chain interactions and anisotropy in SrNi$_{2}$V$_{2}$O$_{8}$ makes it a model system to study the effect of these perturbations on the magnetic ground state as well as on the nature of the field-induced magnetic state. In addition the recent availability of single crystal samples of SrNi$_{2}$V$_{2}$O$_{8}$ allows a much more detailed and accurate investigation to be performed compared to PbNi$_{2}$V$_{2}$O$_{8}$ for which only powder samples are available. The compound SrNi$_{2}$V$_{2}$O$_{8}$ crystallizes in the tetragonal symmetry with space group of {\it I}4$_{1}cd$.\cite{WichmannRCM.23.1} A significant structural feature is that all magnetic Ni$^{2+}$ ions $(3d^8,  S =1)$ are equivalent with arrays of edge-shared NiO$_{6}$ octahedra around the fourfold axis forming screw-chains along the crystallographic $c$-axis. The screw-chains are connected to each other by nonmagnetic VO$_{4}$ (V$^{5+}$; $(3d^0,  S = 0)$  tetrahedral which provide the interchain superexchange interaction pathways in this compound.\cite{BeraPRB.86.024408} 

In this paper, we report the first investigation of the magnetic properties of SrNi$_{2}$V$_{2}$O$_{8}$ using single crystal samples. We have employed static susceptibility, inelastic neutron scattering, high-field magnetization, and ultra-low temperature heat capacity techniques to study the ground state, field-induced magnetic ordering, and the effect of anisotropy. Our results confirm that SrNi$_{2}$V$_{2}$O$_{8}$ has a spin-singlet ground state and a gap between the ground state and first excited triplet states. Anisotropy induced splitting of the triplet states into two modes is found by neutron scattering. The intensity ratio of the modes confirms that the easy axis is along the crystallographic $c$-axis. Field-induced magnetic ordering is found with two critical fields  [${\mu}_{0}${\it H}$_{c}^{\perp c}$= 12.0${\pm}$0.2 T and ${\mu}_{0}${\it H}$_{c}^{\parallel c}$ = 20.8${\pm}$0.5 T at 4.2 K] due to closing of the gaps by Zeeman splitting. The two critical fields are corresponding to the anisotropic Zeeman splitting of the excited triplet branches. 3D long-range magnetic ordering above {\it H}$_{c}$ is evident in the susceptibility, magnetization and heat capacity studies. Substantial inter-chain interactions (which are too weak to induce long-range magnetic ordering at zero field) are also evident and play an important role in inducing the 3D magnetic ordered state above $H_c$. 

\section{\label{sec:exp} EXPERIMENTAL DETAILS}

Single crystals of SrNi$_{2}$V$_{2}$O$_{8}$ were grown by using a traveling solvent floating zone technique at the Crystal Laboratory, Helmholtz Zentrum Berlin f\"ur Materialien und Energie (HZB), Berlin, Germany (details of the technique will be reported elsewhere). The phase purity of the crystals was confirmed by powder X-ray diffraction. A part of each crystal was ground into powder for this study. The back-scattering X-ray Laue patterns revealed the good single crystalline nature of the samples. 

Static (DC) magnetic susceptibility and field dependent (H $\le$ 14 Tesla) magnetization measurements were carried out using a Physical Properties Measurement System (14 T PPMS; Quantum Design) at the Laboratory for Magnetic Measurements, HZB. For these measurements, the crystals were cut into parallelepipeds of dimensions (~3$\times2\times1$ mm$^3$) with edges parallel to the principal crystallographic axes and the data were collected with a magnetic field applied both parallel and perpendicular to the chain direction (crystallographic $c$-axis). The temperature dependent static susceptibility [$\chi(T)$] measurements were performed over a temperature range of 2--550 K under different applied magnetic field of 1--14 Tesla. All measurements were performed in the warming cycle after cooling in zero field. The isothermal magnetization measurements were performed at different temperatures over 1.8--10 K also in the zero field cooled condition. 

The magnetic excitation spectra were investigated using inelastic neutron scattering. The cold neutron triple-axis spectrometer PANDA at FRM-II, Garching, Germany was employed to measure the low energy magnetic excitation spectra. A large cylindrical single crystal of mass $\sim$ 2.5 g (diameter: 6 mm and length: 30 mm) was used for these experiments. The sample was mounted on aluminum sample holders. Double focusing pyrolytic graphite (PG002) monochromator and analyzer were used to select the incident neutron energy and the final energy of the neutrons analyzed in the detector, respectively. Two experiments were performed using fixed final neutron wave vectors of $k_f = 2.57 \angstrom^{-1}$ (13.69 meV) and fixed initial neutron wave vectors $k_i = 1.781 \angstrom^{-1}$ (6.57 meV), respectively. A cooled beryllium filter on the scattered side was used to filter out higher order neutrons for the $k_i = 1.781 \angstrom^{-1}$ setting while a PG filter was used in the case of the $k_f = 2.57 \angstrom^{-1}$ setting. The sample was cooled in a standard closed cycle refrigerator which achieved a base temperature of 3.1 K.

The high-field magnetization curves were recorded using a non-destructive pulsed-field magnet (60 T) with pulse duration of 25 ms at the Hochfeld Magnetlabor Dresden (HLD), Helmholtz Zentrum Dresden Rossendorf, Germany. The magnetization signal was detected by an induction method with a standard pick-up coil system.\cite{SkourskiPRB.83.214420} The data were recorded at different constant temperatures over 1.5--15 K, for both applied field ($H \parallel c$ and $H \perp c$) directions. The data for the empty magnetometer (background) were also recorded at each measuring temperature to get the intrinsic background which was then subtracted from each dataset to get the sample magnetization.

The heat capacity of a small sample with mass of 17.8 mg was measured at the Laboratory for Magnetic Measurements, HZB, with an in-house developed apparatus designed for the relaxation method.\cite{Kiefer-thesis} The data were collected over the temperature range 0.3--100 K and magnetic field range 0--14 T for both field directions ($H \parallel c$ and $H \perp c$). For these measurements, a liquid-He based cryomagnet, equipped with a sorption pumped He$^3$ cryostat insert, was used. 

\section{\label{sec:rnd} RESULTS AND DISCUSSION}
 
\subsection{\label{subsec:chi}Static susceptibility}	

\begin{figure} [bp]
\includegraphics[trim=2cm 4.5cm 3cm 1cm, clip=true, width=90mm]{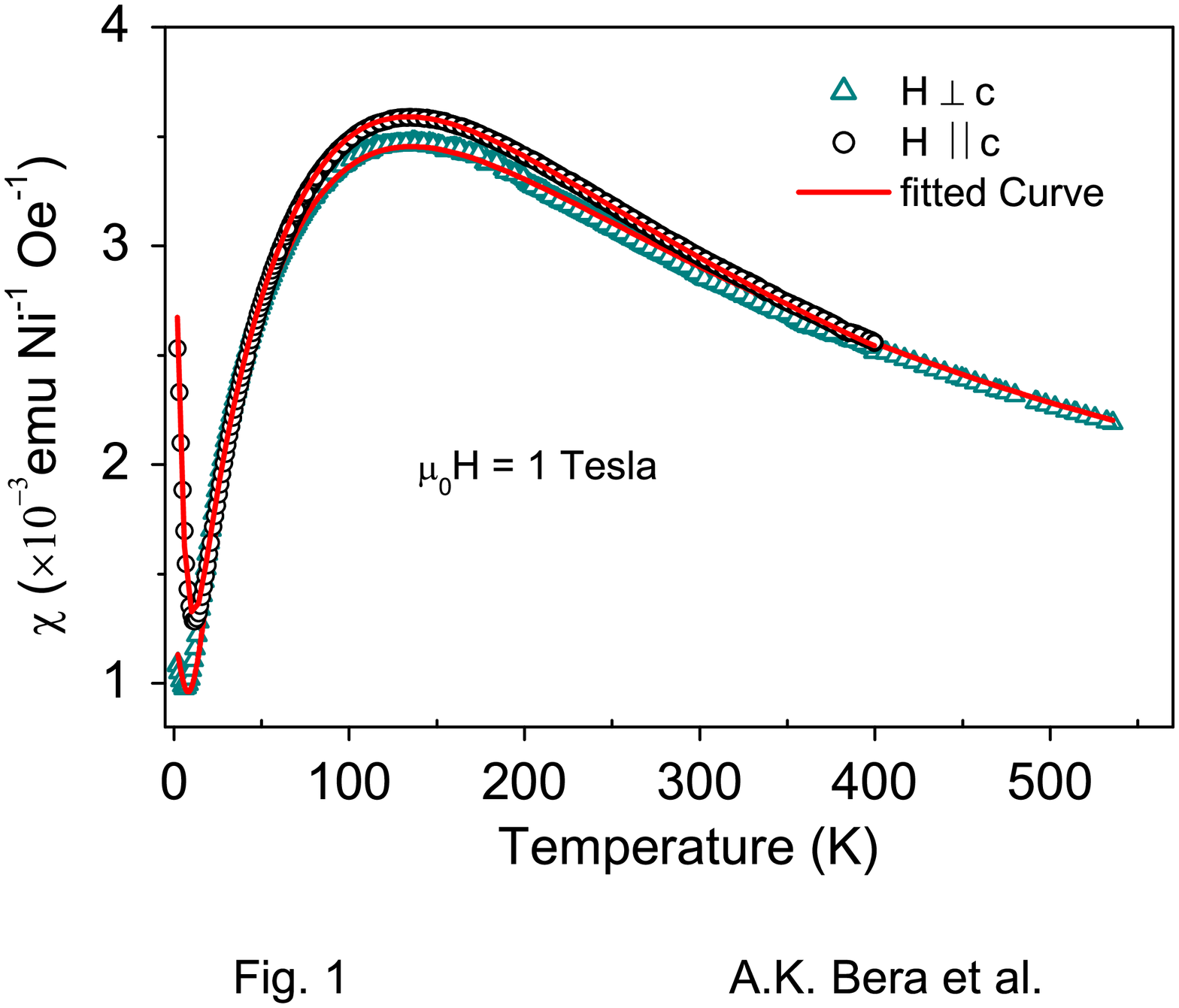}
\caption{\label{fig:susc} (color online)   The $\chi$(T) curves for SrNi$_{2}$V$_{2}$O$_{8}$ under 1 T magnetic field along H $\parallel$ c and H $\perp$ c directions. The red solid lines are the fitted curves according to the Eq.(\ref{eq:one}).}
\end{figure}

The temperature dependent static magnetic susceptibility ($\chi$) curves for the compound SrNi$_{2}$V$_{2}$O$_{8}$ measured under an external applied magnetic field of 1 T both parallel ($c$ axis) and perpendicular ($a$ axis) to the chain direction are shown in Fig. \ref{fig:susc}. With decreasing temperature, the $\chi (T)$ curves show a broad feature over a wide temperature range of 400--15 K (a typical signature of low dimensional magnetic systems due to short-range spin-spin correlations) with maximum around 100 K. Upon lowering the temperature further, the $\chi (T)$ curves decrease down to $\sim$ 10 K and $\sim$ 8 K for $H \parallel c$ and $H \perp c$, respectively, and then show an upturn down to the lowest measured temperature (2 K). The nature of the observed $\chi(T)$ curves is in good agreement with the previously reported $\chi (T)$ curves for powder samples.\cite{PahariPRB.73.012407, BeraPRB.86.024408, HeJPSJ.77.013703} No magnetic transition is found for this compound down to 2 K suggesting spin-liquid behavior.

The observed upturn at low temperatures (below $\sim$ 10 K) is due to the Curie-Weiss contribution arising from natural crystal defects and due to free spins at the ends of finite chains. When a chain breaks due to a lattice defect, two {\it S} = 1/2 free spins are induced at the ends.\cite{Hagiwara.PRL.65.3181, Glarum.PRL.67.1614} This is similar to the impurity-induced (spin-vacancy) magnetization in the {\it S} = 1 chain system by substitution of non-magnetic ions at the magnetic ions site.\cite{MasudaPRB.66.174416, PahariPhysicaB.395.138} In this case the total number of induced free spins is $N$ = 2$\times$$x$$\times$$N_A$, where, $x$ is the fraction of the impurity phase and $N_A$ is Avogadro number. 

Altogether, three contributions are expected to the temperature dependent susceptibility, as,
\begin{eqnarray}
\chi_{obs} = \chi_0 +2x\chi_{imp}+(1-x) \chi_{chain} 
\label{eq:one}
\end{eqnarray}
where, the first temperature independent term $\chi_0$ is due to Van-Vleck paramagnetism as well as diamagnetic core susceptibility. The second term is due to impurities which are responsible for the low temperature upturn, and has the form
\begin{eqnarray}
\chi_{imp} = \frac{N_A\mu_B^2 g_{imp}^2 S(S+1)}{3k_B (T-\theta_{imp})}
\label{eq:two}
\end{eqnarray}
				
where $\chi_{imp}$ is their Curie-Weiss temperature and other constants have their usual meanings. The pre-factor $x$ in the Eq.(\ref{eq:one}) is the fraction of the impurity phase. The third term, $\chi_{chain}$ is the intrinsic susceptibility arising from the Haldane chain SrNi$_{2}$V$_{2}$O$_{8}$. Recently, Law {\it et. al}.,\cite{LawJPCM.25.065601} predicted a general expression of susceptibility for Heisenberg AFM spin-chains with various spin values from the Pad{\'e} approximation. The general expression for the susceptibility is 
\begin{widetext}
\begin{eqnarray}
\chi_{chain} =  \frac{N_A \mu_B^2 g^2 S(S+1)}{3k_B T}\times\exp(\frac{-\Delta J_{NN}}{k_B T})
\times\frac{1+\sum_{i=1}^m A_i (\frac{J_{NN}}{k_B T})^i}
{1+\sum_{j=1}^n B_j (\frac{J_{NN}}{k_B T})^j} 
\label{eq:three}
\end{eqnarray}
\end{widetext} 
where, the term exp($-\Delta$ $J_{NN}/k_BT$) is for the integer spin chains which have a gap at low temperatures, $\Delta$ is the relative gap between the singlet ground and triplet excited states, and $J_{NN}$ is the nearest-neighbor intra-chain super-exchange interaction. The measured susceptibility data was fitted to Eq.(\ref{eq:one}) and a good agreement was found over the full temperature range for both field directions [Fig. \ref{fig:susc}]. The fitted parameters are given in Table \ref{tab:T1}. For the impurity phase, {\it S} and {\it g} values were considered to be 1/2 and 2, respectively. For SrNi$_{2}$V$_{2}$O$_{8}$, the fitted {\it g}-factor values of $\sim 2.34$  for $H \parallel c$ and $\sim2.13$ for $H \perp c$ are close to the value 2.24 estimated from the ESR study at 100 K.\cite{WangPRB.87.104405} For the fittings, the values of the coefficients $A_i$ and $B_j$ are adopted from ref. \onlinecite{LawJPCM.25.065601}. The fitted value of $J_{NN}/k_B,  = 98.53 \pm 0.06$ K is in good agreement with the value reported for the powder samples $\sim$ 102 K.\cite{PahariPRB.73.012407} The thermal activation energy ($\Delta.J_{NN}/k_B$) is found to be $47.5\pm0.7$ K. 

\begin{table} [bp]
\caption{\label{tab:T1}The values of the parameters derived from the temperature dependent susceptibility data (Fig. \ref{fig:susc}).  }
\begin{ruledtabular}
\begin{tabular}{lcc}
Parameters&H $\parallel$ c&H $\perp$ c\\
\hline
$J_{NN}/k_B$ (K) & \multicolumn{2}{c}{98.53$\pm$0.06 (8.5 meV)}\\
$\Delta${\it J}$_{NN}/k_B$ (K) & \multicolumn{2}{c} {47.5$\pm$0.7}\\
{\it g} & 2.342$\pm$0.001 & 2.1292$\pm$0.0002\\
{\it x}	& 0.0656$\pm$0.0002 & 0.0152$\pm$0.0003\\
$\theta_{imp}$ (K) & $-3.03\pm$0.01 & $-2.98\pm$0.07\\
$\chi_0$ [$\times$ 10$^{-4}$ emu (mol-Ni)$^{-1}$] & 5.65$\pm$0.04	 & 12.08$\pm$0.03\\
\end{tabular}
\end{ruledtabular}
\end{table}

\subsection{\label{subsec:ns}Inelastic neutron scattering}

\begin{figure}
\includegraphics[trim=0.5cm 3.5cm 1.4cm 0cm, clip=true, width=90mm]{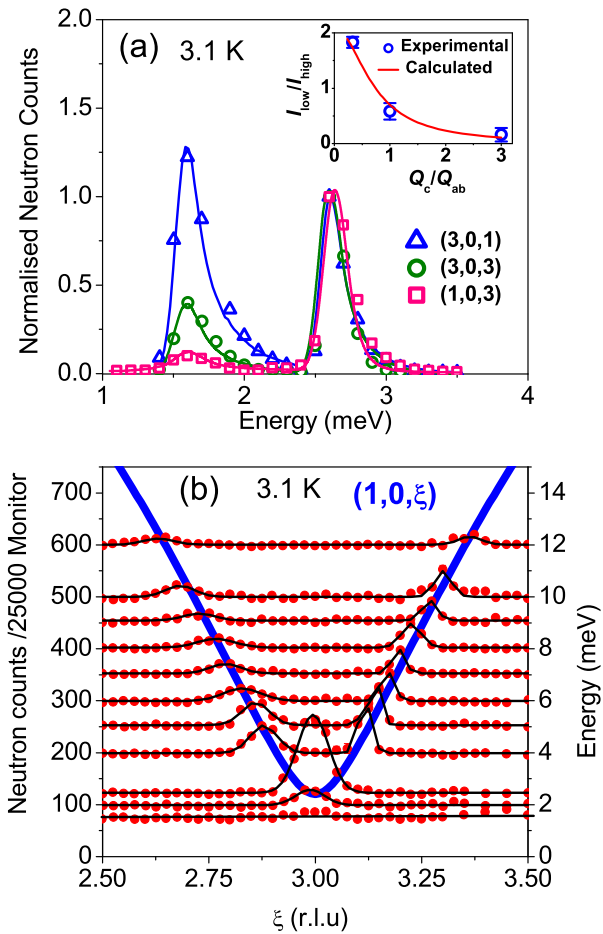}
\caption{\label{fig:ns}  (color online) Neutron scattering results; (a) Constant-wave vector scans for the SrNi$_{2}$V$_{2}$O$_{8}$ compound at the (3,0,1), (3,0,3), and (1,0,3) reciprocal points measured at 3.1 K in the standard three-axis mode. For comparison, the peak intensity of each individual scans is normalized with the intensity of the higher energy peak at $\sim$ 2.6 meV of that spectrum. The solid lines are the fitted curves. The ratio of the intensity of the two peaks ($I_{low}/I_{high}$) as a function of the wave vector component ratio ($Q_c/Q_{ab}$) is shown in the inset by open blue circles. The calculated intensity ratio is also shown by the solid red line. (b) Constant-energy scans at several energy transfers along the chain direction (1,0,$\xi$) at 3.1 K. Vertical (intensity) shifting of the spectra have been done to normalize with the energy transfer. Statistical error bars are smaller than the symbol size. Thin black solid lines through the data points are the least square fit to the observed data. The thick blue solid line is the fitted dispersion curve according to the relation described in the text [Eq.(\ref{eq:five})].}
\end{figure}

In ordered to confirm the gap values and anisotropy in SrNi$_{2}$V$_{2}$O$_{8}$, we have performed, for the first time, the inelastic neutron scattering measurements using a single crystal. Figure \ref{fig:ns} shows the magnetic excitation spectra (constant {\it Q}-scans) measured over 1--3.5 meV at the AFM zone centers (3,0,1), (3,0,3), and (1,0,3) at 3.1 K. These measurements were performed in the fixed incident neutron energy mode with $E_i$ = 6.57 meV ($k_i = 1.781 \angstrom^{-1}$). All excitation spectra are found to be gapped, and reveal two peaks in the energy scan with maxima at $\sim$ 1.6 meV and 2.6 meV, respectively. The observed results are in agreement with Haldane's conjecture for integer spin-chains that they have a singlet ground state and gapped magnon excitations. The presence of two energy gaps for SrNi$_{2}$V$_{2}$O$_{8}$ is due to single-ion anisotropy induced zero-field splitting of the triplet states into two excited branches. To estimate the gap values, we have fitted the observed spectra with Gaussian functions which are convolved with the instrumental resolution [solid lines in Fig. \ref{fig:ns}(a)]. The fitted values of energy gaps are 1.56$\pm$0.1 and 2.58$\pm$0.1 meV for (3,0,1), 1.57$\pm$0.1 and 2.57$\pm$0.1 meV for (3,0,3), and 1.57$\pm$0.1 and 2.60$\pm$0.1 meV for (1,0,3), respectively. We have considered the average values of 1.57 and 2.58 meV for the energy gaps in the rest of the paper. 

The intensity of the low energy peak at $\sim$ 1.57 meV is found to strongly depend on the wave vector direction whereas the intensity of the higher energy peak does not. To make a direct comparison, we normalized the intensity of each individual excitation spectrum with respect to the intensity of the higher energy peak (at $\sim$ 2.58 meV) of that spectrum. A significant decrease in the intensity of the lower energy peak has been found with increasing wave-vector component along the crystallographic $c$-axis or chain direction. To quantify the directional dependence, we have plotted the intensity ratio of the lower energy peak ({$I_{low}$) to the higher energy peak ($I_{high}$) as a function of the wave vector component ratio ($Q_c/Q_{ab}$) in the inset of Fig. \ref{fig:ns}(a). A rapid decrease of the intensity ratio with the increase of the wave vector component along $c$-axis is evident. 

The intensity of neutron magnetic scattering strongly depends on the relative directions of the magnetization vector and the scattering vector which can be used to determine the magnetization of each mode. The directional dependence is given by the factor $[1-(\widehat{Q}.\widehat{\eta})^2$] where, $\widehat{Q}$ is the unit vector along scattering wave vector and $\widehat{\eta}$̂ is the unit vector along the direction of the magnetization. The crystal structure of SrNi$_{2}$V$_{2}$O$_{8}$ is tetragonal with the $c$-axis being the unique axis which suggests that the magnetization easy-axis is either along the $c$-axis ($\widehat{\eta} \parallel c$) or perpendicular to the $c$-axis ($\widehat{\eta} \perp c$).  Assuming that the lower energy mode corresponds to fluctuations along the $c$-axis while the upper mode corresponds to fluctuations in the $ab$-plane and using the above expression, the intensity ratio is calculated as a function of the wave vector component ratio ($Q_c/Q_{ab}$), and shown in the inset of Fig. \ref{fig:ns}(a) by the solid line. Excellent agreement between the experimental and calculated intensity ratios was found confirming that the lower energy mode corresponds to fluctuations along $c$-axis.  This shows that the energy cost for the spins to align along the $c$-axis is less than in the $ab$-plane. Thus neutron scattering confirms that the direction of the magnetic easy-axis is along the crystallographic $c$-axis i.e., along the chain-direction.

The value of the single-ion anisotropy has been estimated below from the size of the splitting of the triplet states. The relation between the energy gaps and the single ion anisotropy was proposed by quantum Monte Carlo simulations,\cite{GolinelliPRB.45.9798} for systems with planar anisotropy ({\it D} $> $ 0), to be 
\begin{eqnarray}
\Delta _+ = \Delta _0+1.41D   \text{		and	}   \Delta _- = \Delta _0-0.57D
\label{eq:four}
\end{eqnarray}
where $\Delta _+$ and  $\Delta _-$ are the values of the energy gaps of the two split triplet states from the singlet ground state,   $\Delta _0$  is the mean gap value in the absence of the anisotropy, and {\it D} is the single ion anisotropy. For the case of uniaxial anisotropy ({\it D} $<$ 0), as evident for SrNi$_{2}$V$_{2}$O$_{8}$, the energy labels must be swapped. Now, using the gap values  $\Delta _\parallel$ = 1.57 meV and  $\Delta _\perp$ = 2.58 meV, the values of mean gap and the single ion anisotropy were calculated to be  $\Delta _0$ = 2.29 meV and {\it D} = $-0.51$ meV, respectively.

We have also studied the low-energy magnetic dispersion along the chain direction ($c$-axis) by performing a series of constant {\it E}-scans at several energy transfers. The constant-{\it E} excitation spectra arround (1,0,3), measured at 3.1 K, are shown in Fig. \ref{fig:ns}(b). These measurements were performed in constant final energy mode with $E_f$ = 13.69 meV ($k_f = 2.57 \angstrom^{-1}$). Due to the relaxed resolution in this experimental configuration the splitting of the triplet state due to single-ion anisotropy is not resolved, and only a single mode is observed. We have fitted the observed dispersion pattern with the well-defined dispersion relation proposed for the Haldane spin-chain compounds\cite{SakaguchiJPSJ.65.3025} as
\begin{eqnarray}
[h\omega(q)]^2 = \Delta ^2+\nu^2\sin^2(2\pi q)
\label{eq:five}
\end{eqnarray}
 where, $\Delta$ is the gap energy, $\nu$ is the spin-wave velocity = 2.49{\it J},\cite{ZheludevPRB.62.8921} {\it J} = AFM super-exchange interaction along the chain-direction and {\it q} is the momentum transfer along the chain direction. The crystal structure of the screw spin-chain compound SrNi$_{2}$V$_{2}$O$_{8}$ contains four equivalent Ni$^{2+}$ ions/spins per unit cell along the chain direction ($c$-axis).\cite{BeraPRB.86.024408} Therefore, the relation between the reduced wave vector transfer along the chain axis and the actual wave vector transfer $Q (= ha^*+kb^*+lc^*)$ is $q = Q_c/4 = l/4$. For the fitting, we have fixed the gap value to the mean gap value of $\Delta_0$ = 2.29 meV (obtained from the two gap values 1.57 and 2.58 meV in the high resolution configuration) expected in the absence of single-ion anisotropy. The fit yields the exchange value {\it J} = 8.9$\pm$0.1 meV which is in good agreement with the value derived from the susceptibility data (Fig. \ref{fig:susc}). 

It may be noted that the mean gap value $\Delta_0$ (2.29 meV) is much smaller than that predicted theoretically for an isolated Haldane chain of $0.41J \sim (0.41\times 8.9$ meV) = 3.65 meV.\cite{WhitePRB.77.134437} The lower value of the mean gap energy suggests the presence of interchain interactions. The presence of interchain coupling suppresses the Haldane gap, and can induce 3D long-range magnetic order upon achieving a critical value. It may be noted here that the substantial inter-chain interaction in the quasi-1D spin-1 chain compound CsNiCl$_3$ results in 3D long-range magnetic ordering below 4.8 K.\cite{MorraPRB.38.543, KadowakiJPSJ.56.751} It is also worth mentioning that from the powder inelastic neutron scattering study on SrNi$_{2}$V$_{2}$O$_{8}$ by Zheludev {\it et. al.}, \cite{ZheludevPRB.62.8921} have estimated an interchain interaction ($J_\perp$) of $\sim$ 0.18 meV. To investigate the inter-chain interactions in detail, measurements of the dispersions perpendicular to the chain-direction (in the $ab$-plane) are required and are in preparation.

\subsection{\label{subsec:hfm}High Field Magnetization}

In order to the study the field induced magnetic ordering and the magnetic anisotropy in further detail; we have performed high-field magnetization measurements using a pulsed magnet. The field-dependent magnetization curves at $T = 4.2$ K for the field parallel and perpendicular to the chain direction (crystallographic $c$-axis) are depicted in Fig. \ref{fig:mhhf}. Both magnetization curves show an abrupt change of slope at critical fields $\mu_0H_c^{\perp c} = 12.0{\pm}$0.2 T  and $\mu_0H_c^{\parallel c} = 20.8{\pm}$0.5 T confirming the field induced closing of the Haldane gap by Zeeman splitting. The critical fields were determined from the derivative of the magnetization with respect to the applied field [discussed later in Fig. 7(a) and 7 (b)]. The presence of two critical fields corresponds to two energy gaps between the singlet ground state and the triplet excited states as found in the neutron scattering study (Fig. \ref{fig:ns}). This confirms again the presence of single-ion anisotropy which removes the degeneracy of the triplet states partially by splitting it into two levels. The higher value of the critical field for $H \parallel c$ as compared to that for $H \perp c$ reveals that the anisotropy is uniaxial ($D < 0$}) which is consistent with the neutron scattering results. 

\begin{figure}
\includegraphics[trim=0.5cm 2.5cm 1.4cm 0cm, clip=true, width=90mm]{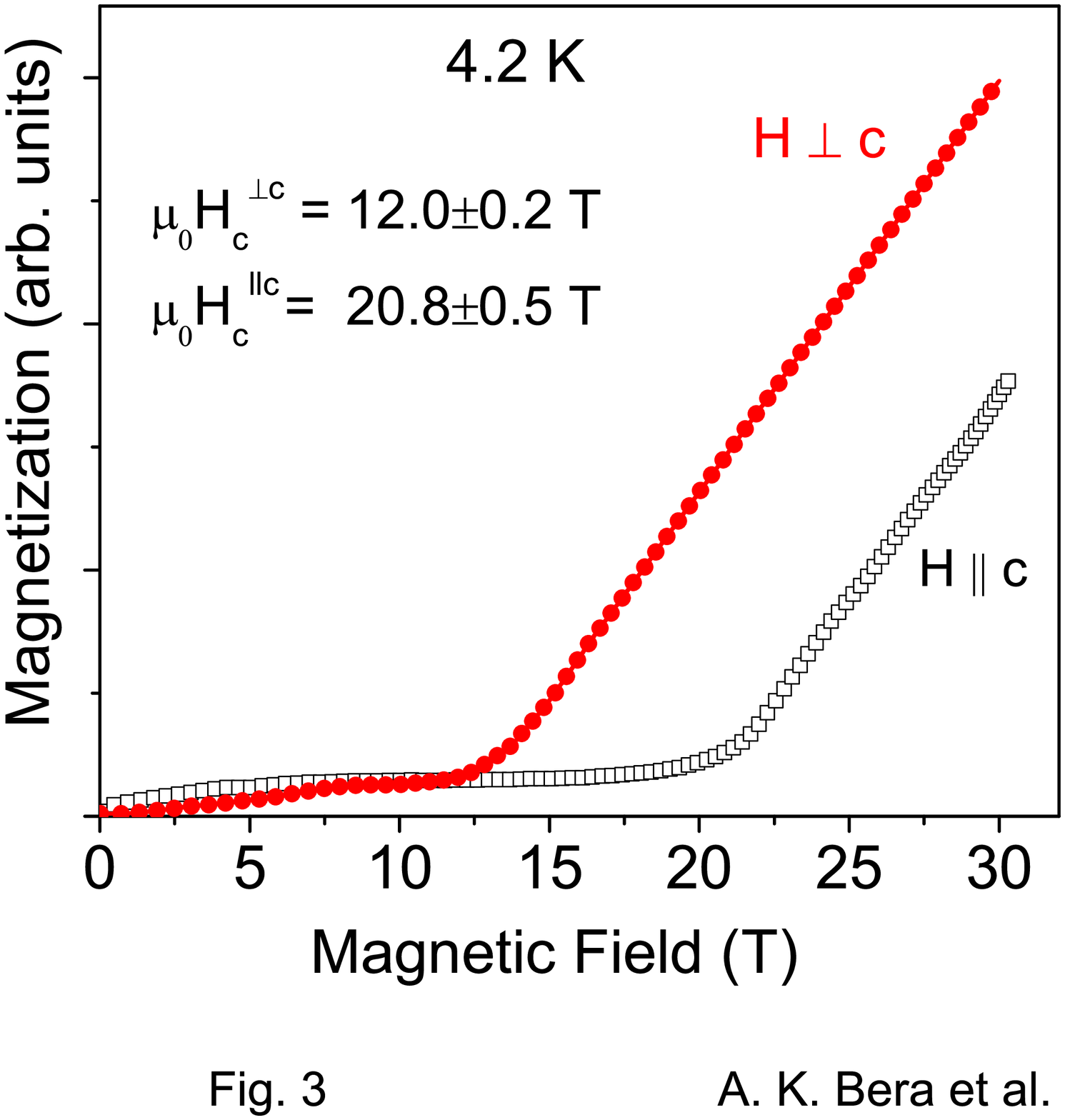}
\caption{\label{fig:mhhf}  (color online) The magnetization vs. magnetic field curves of the compound SrNi$_{2}$V$_{2}$O$_{8}$ for both $H \parallel c$ and $H \perp c$ measured using a pulsed magnet at 4.2 K. The critical fields were determined from the derivative of the magnetization with respective to the applied field where a step appears at the transition field.}
\end{figure}

\begin{figure}
\includegraphics[trim=2.8cm 1.5cm 11.4cm 0cm, clip=true, width=90mm]{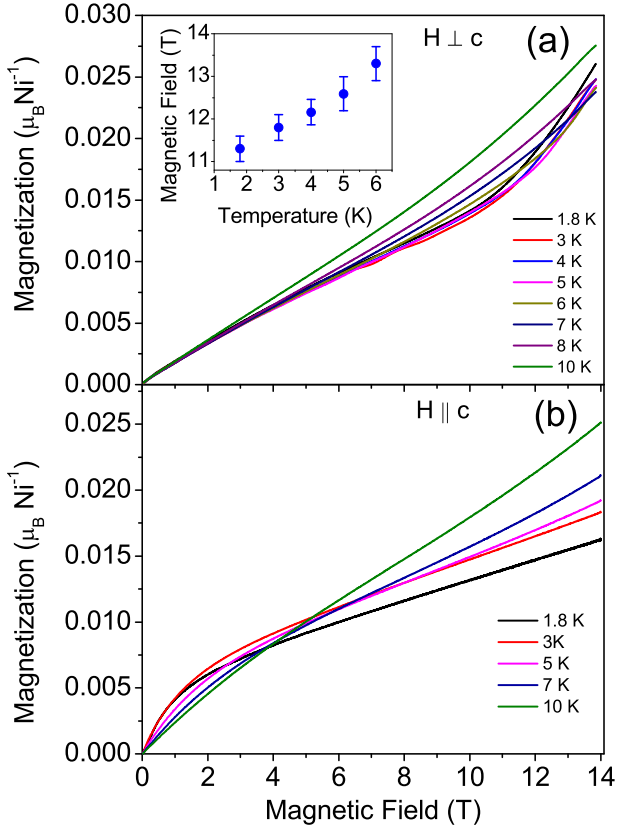}
\caption{\label{fig:mhppms}  (color online) The isothermal magnetization curves as function of magnetic field for SrNi$_{2}$V$_{2}$O$_{8}$ for (a) $H \perp c$ and (b) $H \parallel c$. Inset of (a) shows the temperature dependence of the critical field.}
\end{figure}

For a comparison, we have also measured the isothermal magnetization curves using a PPMS at several temperatures up to 14 T. Such magnetization curves are depicted in Fig. \ref{fig:mhppms}(a) and \ref{fig:mhppms}(b) for the $H \perp c$ and $H \parallel c$, respectively. In agreement with the pulse field magnetization, an anomaly in the magnetization curve has been found at $\sim$ 12.1 T at 4 K for the $H \perp c$. With increasing temperature, the critical field for the magnetic ordering increases from $\sim$ 11.3 T at 1.8 K and becomes higher than 14 T above 6 K [inset of Fig. \ref{fig:mhppms}(a)], and therefore beyond the maximum available magnetic field of the PPMS (14 T). It is also observed that the robustness of the transition decreases with increasing temperature. For the field parallel to the chain direction ($H \parallel c$), as expected, no such transition has been observed up to 14 T [Fig. \ref{fig:mhppms}(b)]. 

Apart from the field induced magnetic transitions at higher fields, a difference between low field magnetizations for $H \parallel c$ and $H \perp c$ is evident at low temperatures. For $H \parallel c$, the magnetization first increases rapidly up to $H \sim 2$ T, though the value of the magnetization is very small $\sim$ 0.005 $\mu_B$ Ni$^{-1}$, and then shows a linear field dependence up to critical field $\sim 20.8$ T.  This type of low field signature in the magnetization curve was reported for the similar compound Pb(Ni$_{1-x}$Mg$_x$)$_2$V$_2$O$_8$ with impurity substitution and was explained due to the disappearance of the impurity induced weak AFM ordering.\cite{MasudaPRB.66.174416} On the other hand, for $H \perp c$, an almost linear field dependence has been observed up to critical field $\sim 12$ T. The differences in the low field magnetization may be due to the different contributions of the impurity magnetization along different applied field directions. This is in agreement with the temperature dependent susceptibility data (Fig. \ref{fig:susc}) where a relatively large Curie-Weiss contribution has been found for $H \parallel c$ as compared to that for the $H \perp c$. 

Now we calculate the values of the energy gaps from the critical magnetic field values to get more details on the nature of magnetic ordering and the effect of anisotropy. In the exchange approximation, the critical field value $H_c$ and the energy gap ($\Delta$) are related by a simple relation $g\mu_B\mu_0H_c = \Delta$. In the presence of the single-ion anisotropy, the degeneracy of the triplet states is removed by zero-field splitting and the field dependences of the gaps become more complex. The effect of the single-ion anisotropy on the Haldane state was approximated theoretically by several methods, such as (i) by a perturbative approach which corresponds to the fermion model of the critical behavior,\cite{GolinelliJPCM.5.7847, RegnaultJPCM.5.L677} (ii) the exact diagonalization for finite chains,\cite{GolinelliJPCM.5.7847} and (iii) by the bosonic version of macroscopic field-theory methods which corresponds to BEC behavior.\cite{AffleckPRB.46.9002} In the magnetic field range well below $H_c$, all models result in the same energy levels. However, they predict different results close to the critical field.  The boson model includes an interchain coupling of nearly the critical value, to cause 3D magnetic order. 

The perturbative approach was employed to estimate the gap energies for the isostructural Haldane-chain compound PbNi$_{2}$V$_{2}$O$_{8}$ which also has an uniaxial-anisotropy.\cite{UchiyamaPRL.83.632} For the uniaxial case, the perturbative approach yields critical fields by the extrapolation of the field up to the closing of the spin-gap. For the field orientation parallel and perpendicular to the crystal axis the expressions are\cite{GolinelliJPCM.5.7847}
\begin{eqnarray}
\Delta_\perp = g\mu_B\mu_0H_c^{\parallel c}  \text{		and	}   \sqrt{\Delta _\parallel\Delta_\perp} = g\mu_B\mu_0H_c^{\perp c}
\label{eq:six}
\end{eqnarray}
where $\Delta_\perp$ is the gap of the $S_z = \pm1$ triplet components with the momentum $q = \pi$, and $\Delta_\parallel$ is the gap of the $S_z = 0$ component. By considering the perturbation approach, from the critical field values, the gap energies are calculated to be $\Delta_\parallel$ = 0.9 meV and $\Delta_\perp$ = 2.69 meV, respectively, considering the g value of 2.24 as estimated from ESR study.\cite{WangPRB.87.104405} The energy gap value of $\Delta_\parallel$ (= 0.9 meV) estimated from perturbation approach is significantly different from the value (1.57 meV) found in the neutron scattering study [Fig. \ref{fig:ns}(a)]. The energy gap values found in the neutron scattering study are correct since this method gives a direct microscopic measurement of the gaps at zero applied magnetic field. Therefore, the extrapolation of the perturbative approach to high fields (in the critical field limit) may not be valid for the present results. 

The exact diagonalization of the finite chain problem,\cite{GolinelliJPCM.5.7847} and the bosonic version of the macroscopic field theory treatment\cite{AffleckPRB.46.9002} give different results for the energy gaps. According to the boson model the gap energies are related to the critical magnetic fields as 
\begin{eqnarray}
\Delta_\perp = g\mu_B\mu_0H_c^{\parallel c}  \text{		and	}   \Delta _\parallel = g\mu_B\mu_0H_c^{\perp c}
\label{eq:seven}
\end{eqnarray}
By using the relations [Eq. (\ref{eq:seven}); bosonic model], the values of the energy gaps are calculated to be $\Delta _\parallel$ = 1.56 meV and $\Delta _\perp$ = 2.69 meV, respectively, from the critical field values ${\mu}_{0}${\it H}$_{c}^{\perp c}$= 12.0${\pm}$0.2 T and ${\mu}_{0}${\it H}$_{c}^{\parallel c}$ = 20.8${\pm}$0.5 T at 4.2 K. These values of the gap energies match well with $\Delta _\parallel$ = 1.57 meV and $\Delta _\perp$ = 2.58 meV at 3.1 K obtained from the inelastic neutron scattering study. Therefore, the relation between critical fields and gap energies agree with the bosonic model, not the perturbative model based on the fermion theory, implying 3D magnetic ordering above the critical field ($H > H_c$) due to the condensation of weakly interacting Bosons. For the iso-structural compound PbNi$_{2}$V$_{2}$O$_{8}$, a recent ESR study also proposed that the field dependence of the gap energies are according to macroscopic boson model\cite{AffleckPRB.46.9002} and macroscopic theory\cite{FarutinJEPT.104.751} (which is similar to the boson model) and not to the perturbative approach or fermionic model.\cite{SmirnovPRB.77.100401} The 3D magnetic order at $H > H_c$, as evident in the present study, is in contrast to the field induced gapless 1D Luttinger liquid state for an isolated Haldane chain. The presence of inter-chain interactions in SrNi$_{2}$V$_{2}$O$_{8}$ may thus play an important role to induce 3D magnetic ordering above the critical field $H_c$.

\subsection{\label{subsec:hfm}High field Susceptibility}

\begin{figure}
\includegraphics[trim=3.5cm 2.5cm 4.0cm 1cm, clip=true, width=90mm]{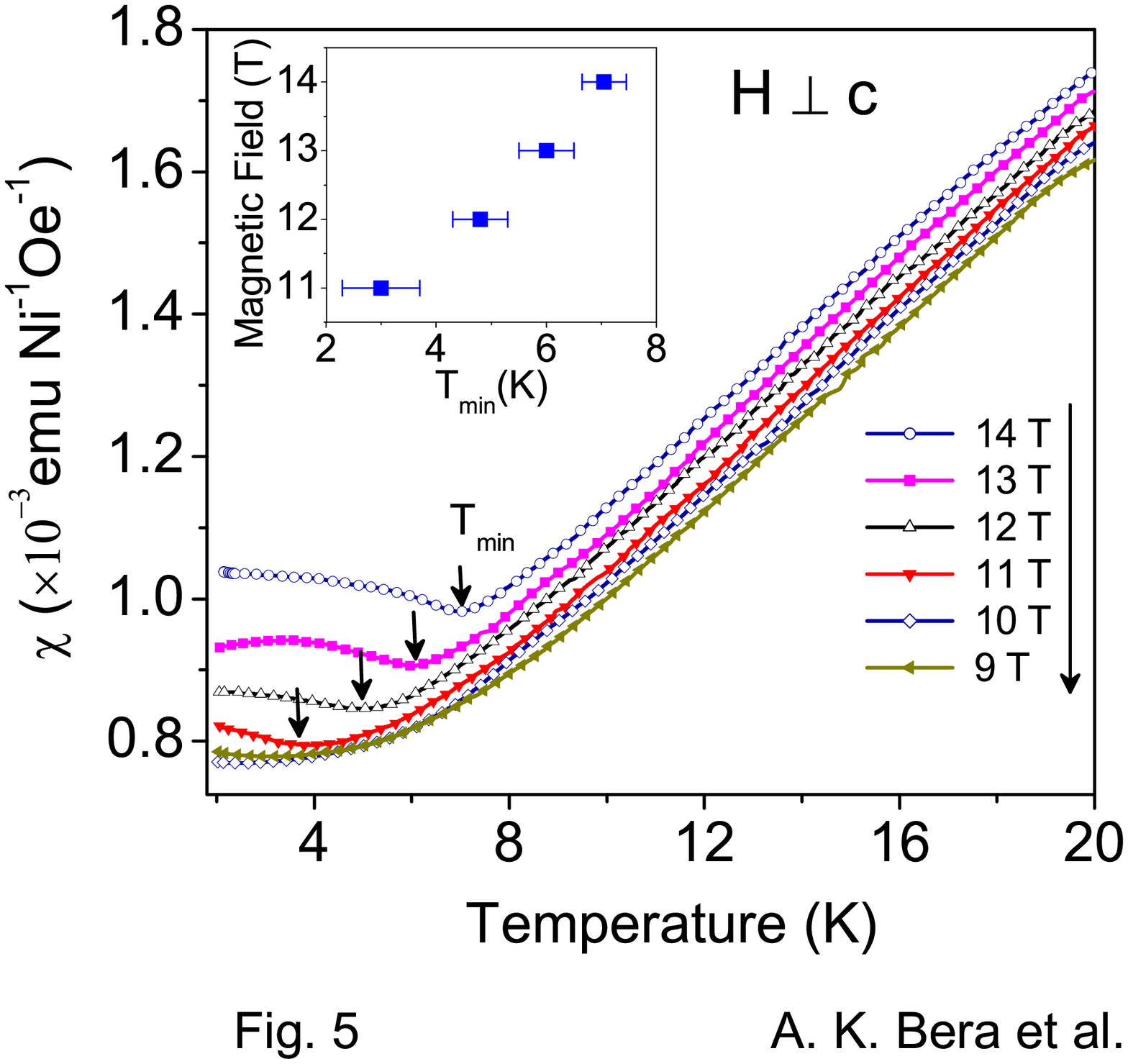}
\caption{\label{fig:chippms}  (color online) The low temperature $\chi(T)$ curves for SrNi$_{2}$V$_{2}$O$_{8}$ measured under magnetic fields of 9, 10, 11, 12, 13, and 14 T along the $H \perp c$ direction. Arrows indicate the temperature corresponding to the susceptibility minimum. Inset shows the magnetic field dependence of the $T_{min}$.}
\end{figure}

To understand the nature of the field induced magnetic ordering further, we have measured temperature dependent dc-susceptibility for $H \perp c$ up to magnetic fields of 14 T using a PPMS. Such susceptibility curves [$\chi(T)$] over 2--20 K are shown in Fig. \ref{fig:chippms}. An anomaly (cusp-like minimum) in the susceptibility has been observed at and above 11 T. Moreover, below the temperature of the anomaly, the susceptibility curve shows a convex behavior. The temperature of the anomaly ($T_{min}$) has been found to increase systematically from $\sim$ 3 K to $\sim$ 7 K with increasing applied magnetic field from 11 T to 14 T, respectively (inset of Fig. \ref{fig:chippms}). Similar magnetization curves (cusp-like minima and followed by the convex type temperature dependence) for $H > H_c$ were reported for the iso-structural compound PbNi$_{2}$V$_{2}$O$_{8}$ and was explained on the basis of BEC of magnons which corresponds to a 3D magnetic ordering.\cite{TsujiiPRB.72.104402} The $T_{min}$ was considered to be the magnetic ordering temperature. As for the BEC theory the increase of the susceptibility below $T_{min}$ is due to the increase in the number of condensed magnons as temperature is reduced.\cite{NikuniPRL.84.5868} However, the presence of anisotropy in SrNi$_{2}$V$_{2}$O$_{8}$ may change the physics of the condensed state away from the BEC, which is evident in our critical exponent study (discussed later).

\subsection{\label{subsec:hfm}Heat capacity}

\begin{figure}
\includegraphics[trim=2cm 2.5cm 12.5cm 0cm, clip=true, width=90mm]{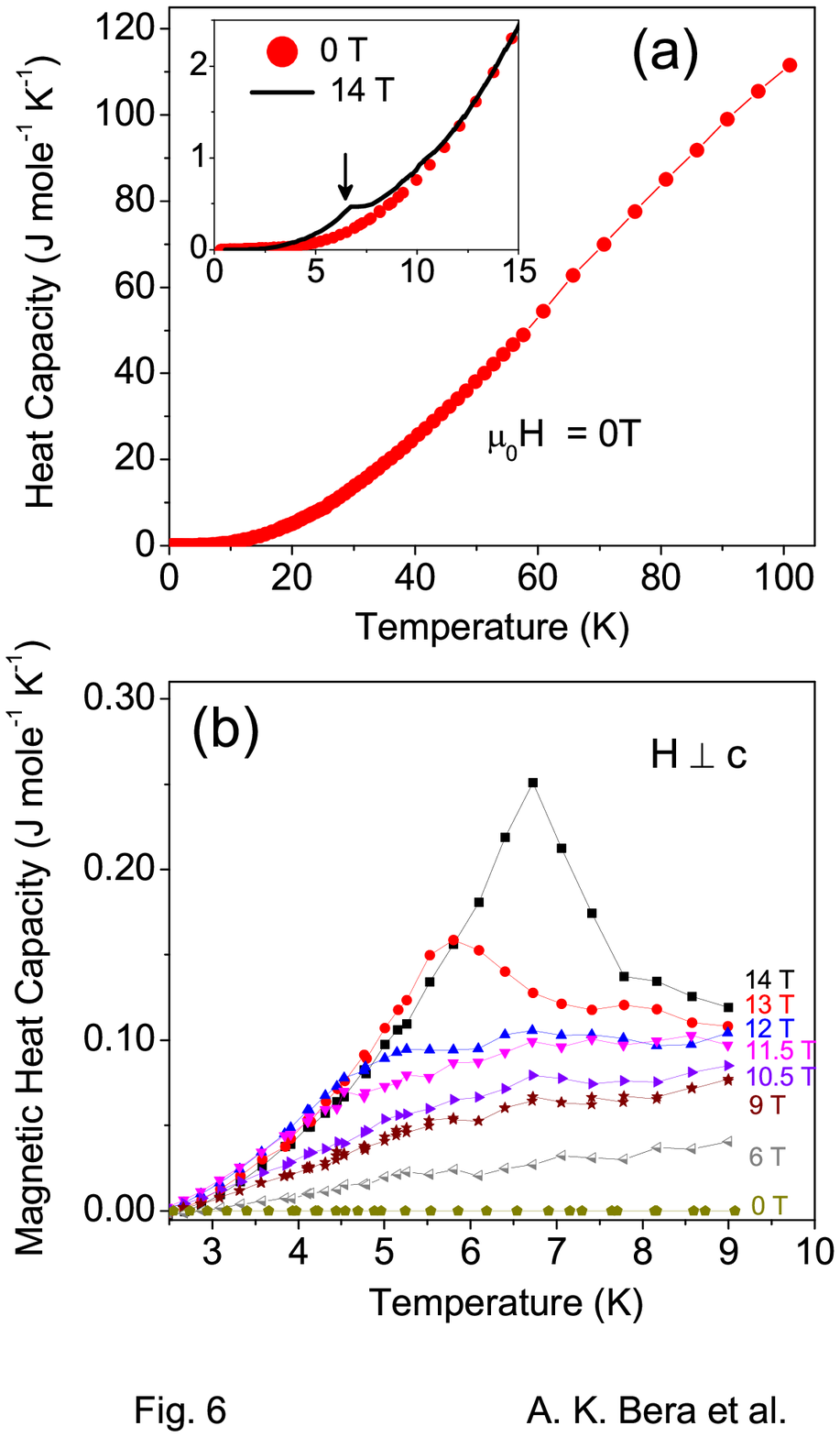}
\caption{\label{fig:cv}  (color online) (a) Temperature dependence of the total heat capacity for SrNi$_{2}$V$_{2}$O$_{8}$ under zero magnetic field. Inset shows the heat capacities measured under 0 T and 14 T over the low temperature region. Arrow shows the transition temperature under 14 T. (b) The temperature dependent field induced magnetic heat capacity, obtained after subtraction of zero field heat capacity [$C_p(H) - C_p(0)$], measured under different magnetic fields from 0--14 T.}
\end{figure}
In order to investigate further the nature of the field-induced magnetic state, we have carried out a heat capacity study on a single crystal. It is generally believed that the specific heat is one of the most powerful measurements to confirm long-range magnetic ordering. The temperature dependent heat capacity curve, measured in zero magnetic field is shown in Fig \ref{fig:cv}(a). No peak in the heat capacity corresponding to the magnetic transition has been found down to 300 mK confirming the singlet ground state. The inset of Fig. \ref{fig:cv}(a) compares the heat capacity vs. temperature curves measured under zero and 14 T magnetic fields ($H \perp c$) over the low temperature region. For the 14 T curve, a sharp peak at $\sim$ 6.75 K has been observed. This peak corresponds to the phase transition to the long-range magnetic ordered state. To study the nature of the peak in more details, we have extracted the field induced magnetic heat capacity ($C_m$) curves by subtracting the zero-field heat capacity [Fig. \ref{fig:cv}(b)]. For the 14 T magnetic heat capacity curve, the pronounced $\lambda$-like anomaly at $\sim$ 6.75 K confirms the 3D long-range AFM phase transition. It is also evident from the weak peak that the change of the magnetic entropy during this magnetic transition is quite small, indicating large quantum fluctuations even in the long-range ordered state.

\subsection{\label{subsec:hfm}Magnetic Phase diagram}

\begin{figure}
\includegraphics[trim=4cm 8cm 13cm 0.5cm, clip=true, width=90mm]{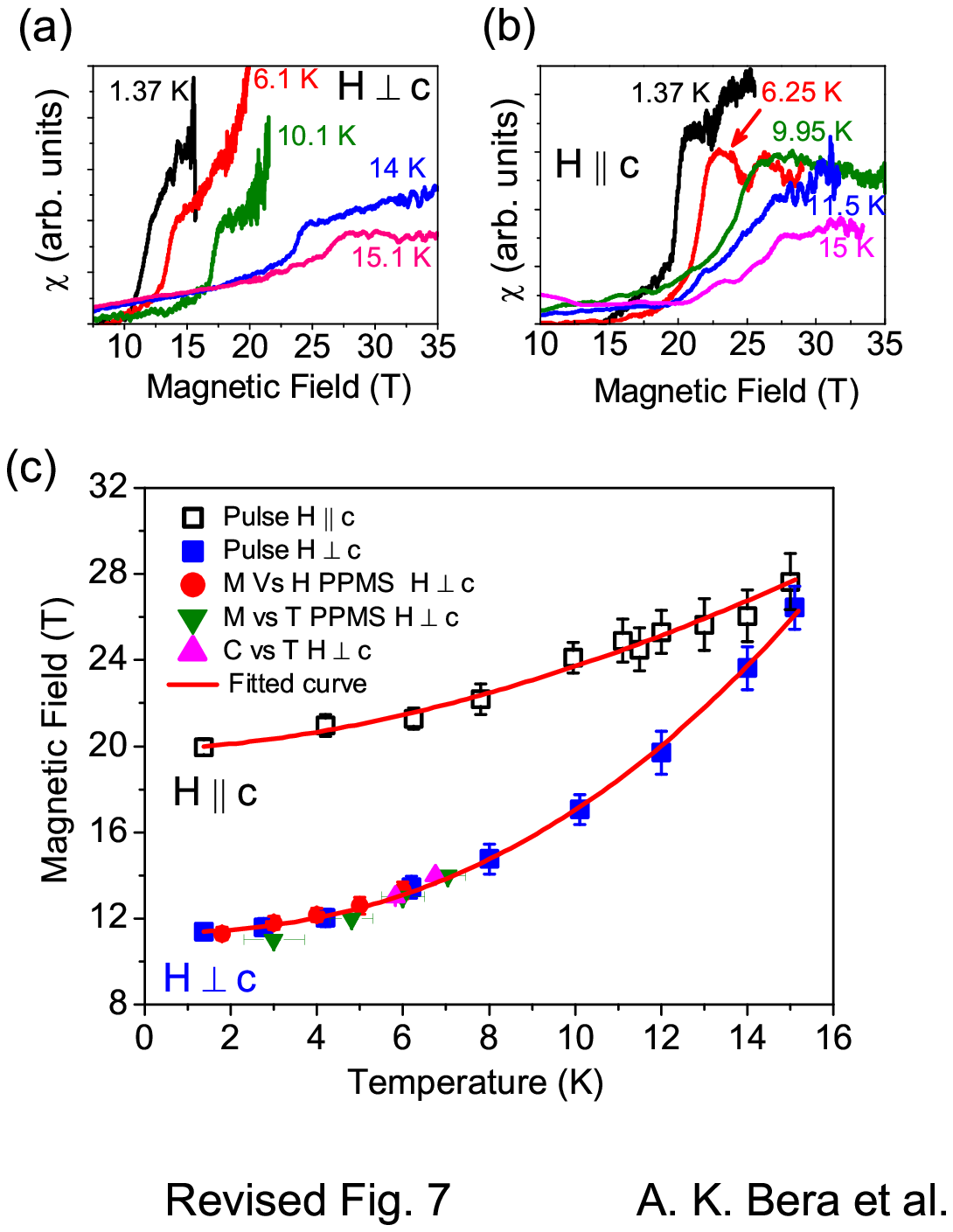}
\caption{\label{fig:pd}  (color online) The magnetic field dependence of the susceptibility ($\chi = dM/dH$), for (a) $H \parallel c$ and (b) $H \perp c$ showing the field induced magnetic transition. (c) The phase diagram showing the critical magnetic field vs. ordering temperature ($T_N$) for SrNi$_{2}$V$_{2}$O$_{8}$. The transition temperatures determined from different measurements are shown with different symbol. The solid lines are fitted curves according to the Eq. (\ref{eq:eight}).}
\end{figure}

BEC is characterized by the critical properties of magnets in the vicinity of the field induced phase transition. Close to the phase transition point or quantum critical point, the phase boundary $T_c(H)$ follows a power law, $T_c \propto (H - H_c)^\phi$ with a universal critical exponent of $\phi = 2/3$ for three dimensional systems.\cite{GiamarchiNaturePhy.4.198} To verify the applicability of the BEC in the case of SrNi$_{2}$V$_{2}$O$_{8}$, we study below the phase diagram. Figures \ref{fig:pd}(a) and \ref{fig:pd}(b) show the susceptibility ($\chi = dM/dH$), as a function of applied magnetic field (pulsed) at several temperatures for $H \parallel c$ and $H \perp c$, respectively. The step in the susceptibility corresponds to the field induced magnetic transition. With increasing temperature, the magnitude of the step in the $\chi(H)$ curve, which indicates the robustness of the magnetic transition, decreases and becomes too weak to measure above $\sim$ 15 K. The magnetic field and temperature phase diagram has been deduced from such susceptibility curves, measured at several constant temperatures. The obtained phase diagram is depicted in Fig. \ref{fig:pd}c).  For both directions, the critical field for the magnetic ordering increases with increasing temperature. The phase boundaries are fitted with the power law of the form

\begin{eqnarray}
T_N = A(H - H_c)^\phi  
\label{eq:eight}
\end{eqnarray}

where $A$ is a proportionality constant. The fitted values of $A$ and $\phi$ are listed in the Table \ref{tab:T2}. The fitted value of $\phi = 0.57\pm0.05$ for the $H \parallel c$ is close to the value 2/3 expected for a 3D BEC. For the field direction $H \perp c$, the value of the exponent  $\phi$ is found to be $0.43 \pm 0.01$ which is indeed expected to be different from the BEC exponent value 2/3. A BEC of magnons is a spontaneous breaking of SO(2) symmetry, and can only occur in a tetragonal geometry, with a field applied along the unique axis which in the present case corresponds to the field applied along the {\it c}-axis. For any other direction of the applied field, the transition is in the Ising universality class and the expected value of the exponent is 1/2.\cite{zapfPRL.96.077204} The fitted value of the exponent $0.43 \pm 0.01$ for $H \perp c$ is close to the theoretical value.

\begin{table}
\caption{\label{tab:T2}The fitted values of the parameters deduced from the fitting of the Eq. (\ref{eq:eight}) to $T_N$ vs. $H$ curves (Fig. \ref{fig:pd}).  }
\begin{ruledtabular}
\begin{tabular}{cccc}
Field direction & $A$ & $\phi$ & $H_c$\\
\hline
$H \parallel c$ & 4.6$\pm$0.5 & 0.57$\pm$0.05&19.9$\pm$0.2\\
$H \perp c$ & 4.7$\pm$0.1 & 0.43$\pm$0.01 & 11.33$\pm$0.07\\
\end{tabular}
\end{ruledtabular}
\end{table}

In the present study, the high-field magnetization, susceptibility and heat capacity measurements on the compound SrNi$_{2}$V$_{2}$O$_{8}$ reveal field induced 3D magnetic ordering above $H_c$.  This is in contrast to the field induced crossover from the gapped singlet state to the gapless Tomonaga-Luttinger liquid (TLL) state at finite temperature predicted for isolated Haldane chains.\cite{MaedaPRL.99.057205} Unlike the ordered state in the BEC, in the case of TLL, the condensed state is disordered due to the thermal and quantum fluctuations because of the one-dimensional nature of the magnetism. The interchain couplings in SrNi$_{2}$V$_{2}$O$_{8}$ thus might be playing an important role in the field-induced condensation of the 3D ordering, though they are not sufficiently strong to induce a magnetic ordered state at zero field. It may be stated here that the sharp peak in the magnetic heat capacity curves (corresponding to the 3D magnetic ordering) are observed for $H \ge$ 13 T [Fig. \ref{fig:cv}(b)], although the magnetization study reveals the critical field, $H_c = 11.3$ T, [Table II]. It may also be noted that with increasing temperature, the difference between phase boundaries for the two field directions [$H_c^{(\parallel)} - H_c^{(\perp)}$] decreases significantly from $\sim$ 8.5 T at 1.5 K to $\sim$ 1 T at 15 K, respectively[Fig. \ref{fig:pd}(c)]. A tendency toward the crossover of the phase boundaries may be evident. However, a theoretical study based on exact diagonalization of finite chains predicted that (i) no intersection of two boundary curves in the case of $D < 0$ (Ising like), while (ii) two intersections in the case of $D > 0$ (XY like) are expected.\cite{sakaiJAP.89.7195} This requires further theoretical study on the phase diagram in the presence of both anisotropy and interchain couplings. On the other hand, the difference between the two critical fields is the measure of the anisotropy. The decrease of the difference between the boundaries may therefore, be due to the softening of the effect of anisotropy with increasing temperature. To understand this point, inelastic neutron scattering measurements as a function of temperature and magnetic field are planned.

\section{\label{sec:rnd}SUMMARY AND CONCLUSION}

We have synthesized the first single crystal of the SrNi$_{2}$V$_{2}$O$_{8}$ compound. Results of static susceptibility, high-field magnetization, inelastic neutron scattering and low-temperature heat capacity studies on the single-crystal provide a consistent picture of SrNi$_{2}$V$_{2}$O$_{8}$ as a Haldane chain compound having a non-magnetic spin-singlet ground state and a gap between the singlet and triplet excited states. The intra-chain exchange interaction has been estimated to be $J \sim 8.9\pm0.1$ meV from the inelastic neutron scattering study. Anisotropy induced splitting of the triplet states into two separate modes with minimum energies of 1.57 meV and 2.58 meV has been found at 3.1 K. The value of the single-ion anisotropy $D$ is derived to be $-0.51\pm$0.01 meV. The easy axis is found to be along the crystallographic $c$-axis i.e., along the chain direction. Field induced magnetic ordering has been found with two critical fields ($\mu_0H_c^{\perp c} = 12.0{\pm}$0.2 T  and $\mu_0H_c^{\parallel c} = 20.8{\pm}$0.5 T for $H \perp c$ and $H \parallel c$, respectively, at 4.2 K) due to closing of the gaps by the Zeeman splitting. When compared to the inelastic neutron scattering results, the boson model estimates the correct values of the gap energies from the critical field values, whereas, the fermion model fails in this context, implying a 3D magnetic ordering above the critical field. The susceptibility and heat capacity data above the critical field ($H_c$) confirm the 3D magnetic ordering. Thus, the interchain couplings in SrNi$_{2}$V$_{2}$O$_{8}$, which are evident from the lower value of the mean gap energy (2.29 meV) as compared to that expected theoretically (3.6 meV) for an isolated Haldane chain, should be important for the field induced ordered state.




%

\end{document}